\documentclass[twocolumn,letterpaper]{IEEEAerospaceCLS}  % only supports two-column, letterpaper format

\usepackage[]{graphicx}    % We use this package in this document
\usepackage{amsmath}
\newcommand{\ignore}[1]{}  % {} empty inside = %% comment

\usepackage{float}
\usepackage{booktabs}
\usepackage{multirow}
\usepackage{amsfonts}

\begin{document}

\title{Algorithmic Approaches to Reconfigurable Assembly Systems}

\author{%
Allan Costa\\ 
Department of Computer Science and Engineering \\
Massachusetts Institute of Technology\\
Cambridge, MA\\
allanc@mit.edu
\and 
Amira Abdel-Rahman\\
Center for Bits and Atoms\\
Massachusetts Institute of Technology\\
Cambridge, MA\\
amira.abdel-rahman@cba.mit.edu
\and 
Benjamin Jenett\\
Center for Bits and Atoms\\
Massachusetts Institute of Technology\\
Cambridge, MA\\
bej@mit.edu
\and 
Neil Gershenfeld\\
Center for Bits and Atoms\\
Massachusetts Institute of Technology\\
Cambridge, MA\\
neil.gershenfeld@cba.mit.edu
\and 
Irina Kostitsyna\\
Department of Mathematics and Computer Science\\
TU Eindhoven\\
Eindhoven, Netherlands\\
i.kostitsyna@tue.nl
\and 
Kenneth Cheung\\
Coded Structures Laboratory\\
NASA Ames Research Center\\
Moffett Field, CA\\
kenny@nasa.gov
%%%% IMPORTANT: Use the correct copyright information--IEEE, Crown, or U.S. government. %%%%%
%\thanks{\footnotesize 978-1-5386-6854-2/19/$\$31.00$ \copyright2019 IEEE}              % This creates the copyright info that is the correct 2019 data.
\thanks{{978-1-5386-5541-2/19/\textdollar31.00 \copyright2019 IEEE}}         % Use this copyright notice only if you are employed by the U.S. Government.
%\thanks{{978-1-5386-6854-2/19/$\$31.00$ \copyright2019 Crown}}          % Use this copyright notice only if you are employed by a crown government (e.g., Canada, UK, Australia).
%\thanks{{978-1-5386-6854-2/19/$\$31.00$ \copyright2019 European Union}}    % Use this copyright notice is you are employed by the European Union.
}

\maketitle

\thispagestyle{plain}
\pagestyle{plain}

\maketitle

\thispagestyle{plain}
\pagestyle{plain}

\maketitle
\begin{abstract}
Assembly of large scale structural systems in space is understood as critical to serving applications that cannot be deployed from a single launch. Recent literature proposes the use of discrete modular structures for in-space assembly and relatively small scale robotics that are able to modify and traverse the structure. This paper addresses the algorithmic problems in scaling reconfigurable space structures built through robotic construction, where reconfiguration is defined as the problem of transforming an initial structure into a different goal configuration. We analyze different algorithmic paradigms and present corresponding abstractions and graph formulations, examining specialized algorithms that consider discretized space and time steps. We then discuss fundamental design trades for different computational architectures, such as centralized versus distributed, and present two representative algorithms as concrete examples for comparison. We analyze how those algorithms achieve different objective functions and goals, such as minimization of total distance traveled, maximization of fault-tolerance, or minimization of total time spent in assembly. This is meant to offer an impression of algorithmic constraints on scalability of corresponding structural and robotic design. From this study, a set of recommendations is developed on where and when to use each paradigm, as well as implications for physical robotic and structural system design.
\end{abstract}

%\begin{IEEEkeywords}
%Space Assembly, Reconfiguration, Algorithms, Robotics, Aerostructures
%\end{IEEEkeywords}

% \IEEEpeerreviewmaketitle

\section{Introduction}

%better underline the relevance of the proposed work to space/orbital systems applications, ie: why does it not apply to any assembly task?

Automated, robotic processes for material deposition and manipulation have become central to state-of-the-art material and structural system production. Additive manufacturing can produce hierarchical, architected materials with nanometer-scale features on centimeter scale parts, resulting in novel properties unattainable with traditional engineering materials \cite{schaedler2016architected}. High performance, continuous fiber composite aircraft components can be made by automated tape laying, utilizing large gantries, tooling, and autoclaves \cite{august2014recent}. However, even with the benefits of electronic digital control systems, some inherent limits exist due to the monolithic material system choice, including stochastic error detection and repair \cite{levesque2016inspection}. 

An alternative strategy is based on the assembly of discrete, building-block elements into larger functional structures, using reversible connections to allow for disassembly and reuse. These two properties, modularity and reconfigurability, have been appreciated throughout the history of engineered structures, and can be found in applications ranging from truss units for  infrastructure to LEGO-style consumer products and toys. Modularity can reduce complexity, and reconfigurability can reduce material requirements for a variety of objectives. Combined, these attributes create significant potential for cost saving, adaptability, and scalability. 

These benefits are particularly desirable in risk-averse, mass-constrained applications, such as space structures. In the relatively short history of space systems, we already see application of these principles in space stations and some large instrumentation \cite{belvin2016space}, with the literature suggesting that further development of the approach may be key to building large scale infrastructure for long duration missions \cite{jefferies2017space}. 

\begin{figure}[ht]
\centering
\includegraphics[width=0.4\textwidth]{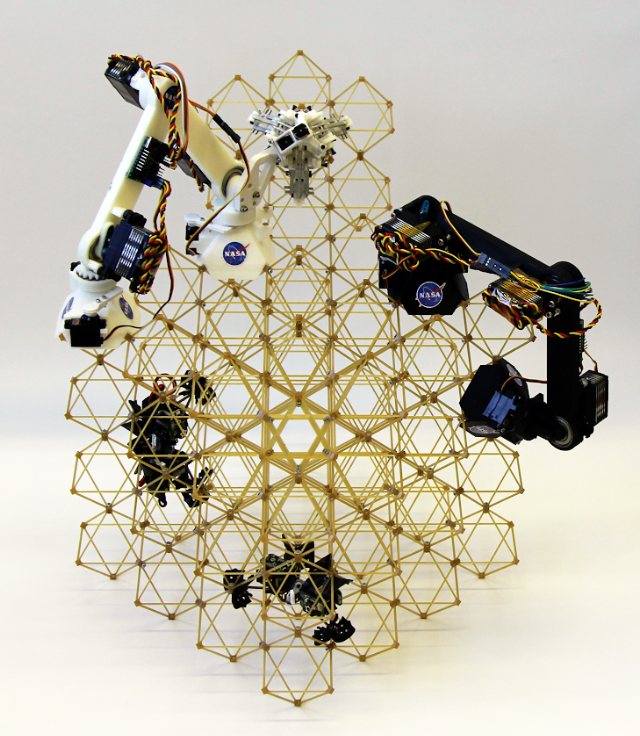}
\caption{Modular reconfigurable 3D lattice structure \protect\cite{gregg2018ultra} and mobile robots \protect\cite{jenett2017bill}, \protect\cite{jenett2017mobile}, showing the small size of the robots relative to the structure that they work on, and the parallel use of multiple robots.}
\label{fig:ARMADAS}
\end{figure}

More recently, modular lattice elements with reversible connections have been assembled into structures with record setting mechanical properties \cite{cheung2013reversibly}, large-scale reconfigurability \cite{jenett2016meso}, and mission-adaptive performance \cite{jenett2017digital}. These lattice elements, also termed voxels, can be mass produced with best-practice manufacturing, ensuring repeatability and affordability \cite{gregg2018ultra}. Further, it has been shown that simple, task-specific robots can be designed for assembly, inspection, and repair \cite{jenett2017bill}\cite{jenett2017mobile}, and that the periodic, structured nature of the material system can be leveraged to reduce robot complexity. NASA's Automated Reconfigurable Mission Adaptive Digital Assembly Systems (ARMADAS) project \cite{armadas} seeks to extend this material-robot system to the assembly of modular, reconfigurable space structures, as shown in Figure \ref{fig:ARMADAS}. 

In this paper, we address the algorithmic challenges faced by a multi-robot system for reconfiguration of these modular, building-block structures. The use of discretized structures simplifies the algorithmic representation, and allows for physical error correction by discrete replacement. It is hoped that the use of discretized structures will also simplify the mechanical robotic assembly problem and demonstrate natural scalability.

\vspace{8mm}

We consider a system in which robots physically occupy a number of voxels' worth of spatial volume whether or not they are staying in one location or moving between adjacent locations. For the purposes of this paper, the robots are spatially constrained to the surfaces of the structure. Both of these constraints have computational implications for path planning and collision-detection. Finally, the structures are assumed to be three dimensional and isotropic in that assembly steps can proceed in any direction, we do assume, however, that the initial and final structure configurations have intersecting regions. The space that robots occupy necessarily changes during reconfiguration steps, and robots are generally presumed to travel between areas that are being disassembled and areas being constructed.

%%%%%%%%%%%%%%%%%%%%%%%%%%%%%%%%%%%%%%%%%%%%%%%%%%%%%%%%%%%%%%%%%%%%%%%%%%%%%%%%%%%%%%%%%%%%%%%%%%%%%%%%%%
\section{Background}
%%%%%%%%%%%%%%%%%%%%%%%%%%%%%%%%%%%%%%%%%%%%%%%%%%%%%%%%%%%%%%%%%%%%%%%%%%%%%%%%%%%%%%%%%%%%%%%%%%%%%%%%%%

Mobile robotic fabrication is a popular topic, with many different approaches. We cover relevant prior art here, including material systems, sensing and positioning, construction and mobility algorithms, and hardware strategies. 

One of the main appeals of mobile robotic fabrication is the ability to build structures larger than the build envelope of static robotic platforms. While large scale additive manufacturing has shown promise \cite{duty2017structure} \cite{khoshnevis2006mega}, the size of the finished product is limited to the extents of the motion system. Mobile additive manufacturing has been tested at various scales \cite{keating2017toward} \cite{zhang2018large}, and a key challenge is maintaining precision over long distances while utilizing a continuous, monolithic material system, ie: deposition of plastic, concrete, or metal. 

An alternative approach is to use a discrete material system, ie: bricks, struts and nodes, or other forms of voxels. One main benefit of a discrete material system is its extensibility. Large scale structures can be decomposed into finite elements, and assuming a mobile robot can successfully place these elements, there should be no limit on scale. This assumption, however, relies on sensing, metrology, and localization. 

Some systems use a global positioning system, such as a Vicon motion tracking system, to determine the location of the mobile robots and the state of the built object. This has been employed for aerial drone construction of brick \cite{augugliaro2014flight}  and strut and node systems \cite{lindsey2012construction}, as well as wheeled, multi-DoF arms for coordinated assembly \cite{dogar2015multi}. While this allows for arbitrary construction within the bounds of the global sensing system, it is ultimately limited to these bounds. 

Other mobile robots leverage the modularity of the discrete construction system to achieve global precision from local metrology. Insect-inspired, wheeled robots can place passive bricks \cite{werfel2014designing}, or inchworm style platforms can traverse and manipulate active bricks \cite{terada2008automatic}. Truss-based structures offer improved structural performance \cite{nigl2013autonomous}. Our work builds on this approach, combining novel material properties for space applications.

We will now look at how mobile, discrete robotic construction systems use algorithms to address two main challenges: material deposition, or build, sequencing and robot path planning.

Within build sequencing, we can differentiate between local and global rules. Local rules describe where a block can or cannot be placed based on the configuration of neighboring blocks. The TERMES project, by Werfel et al, exemplifies this approach. Here, local rules are set up to avoid configurations where a subsequent block cannot be placed \cite{werfel2008three}, or where a robot can not traverse the structure based on limited movements, ie: cliffs or canyons \cite{werfel2011distributed}. Globally, we seek to describe a desired geometry output in a language that can be executed by the robots. In TERMES, this was accomplished using an offline compiler which determined a “"structpath"”, a one-way road which robots would enter and exit, depositing blocks, to achieve the final structure \cite{werfel2014designing}. The Automatic Modular Assembly System (AMAS) applied both discrete and continuous gradient fields to simulate robots being guided from a material supply zone to a build front \cite{terada2006modular}.  

Path planning deals more directly with the motions of the robot. Objectives include decreasing total build time, optimizing multi-robot construction, and avoiding collision. Here we look at two different approaches to the system control architectures: centralized and distributed.

Centralized control architectures can be vulnerable to single points of failure, but also can achieve optimality more easily due to precomputing of a solution. This works for a relatively small number of robots with a relatively small number of parts, and has been implemented in several projects. 

One distributed approach is biologically inspired by a variety of ant species that use stigmergy to deposit and follow pheromone trails \cite{goss1989self}. These ants do not have a global view of their final destination, they only have a local view of the pheromone levels at each intersection, and by making small local decisions, they are able to achieve complex global goals. In the distributed model, we see use of gradient fields, or of a guided path, with local rules.

Lastly, in hardware, we see a number of options that relate to algorithm strategies. In AMAS, it is assumed that blocks can communicate \cite{terada2008automatic}. This requires both mechanical and electrical interconnects, which increases block complexity, but provides robust connectivity and avoid the need for wireless communication. In TERMES, blocks are passive, but markings are used by IR sensors on the robot for localization \cite{werfel2006extended}. Typically, real-world experiments occur in a 1g environment, as opposed to microgravity. This affects how robots interact with the structure. TERMES relies on gravity, with continuous wheeled motion providing lateral motion, and the ability to climb up or down one block level. This simplifies hardware design, but results in a number of algorithmic constraints, such as the inability to climb vertical faces of a structure. Conversely, AMAS is designed to lock into any given face of the cubic blocks, which would allow for full 3D locomotion, though this was never fully physically implemented \cite{terada2008automatic}. 

This paper is one of the very few papers that presents different path planning algorithms for the same system to compare the performance and optimality of each approach. The use of passive building blocks ensures the scalibity of the structures to build large and relatively inexpensive structurally sound systems. The algorithmic approaches discussed are not only able to build a larger family of shapes than similar robotic construction approaches  \cite{werfel2006extended}, it is one of the first to discuss algorithms for the reconfiguration (parallel dissassembly and assembly) of structures from one arbitrary shape to another.

\section{Model}

We now describe the approach based on a discrete lattice material system and mobile robots. A key attribute of the system is the discretization of space by the use of finite voxels or building blocks. We also make use of a strong assumption of a reliable robot synchronization, whereby time is taken as discretized in time steps for the entire system. 

We represent a (discrete) position in space with an integer vector and a boolean value that indicates the existence of a voxel in that position. We refer to the set of space cells that contain voxels as a \textit{lattice}. We refer to the state of the lattice and of the robots at a given time step as a \textit{configuration}, which we represent as $\Gamma(\cdot)$. We then denote $\Gamma(\vec{r})$ as the indicator variable for the existence of a voxel at position $\vec r$. 

The example geometry that we employ here, for clarity, is a simple cubic arrangement of voxels. Each voxel has exactly six faces in which robots can possibly step. A face is said to be \textit{exposed} if there is no voxel in the adjacent position. Robots operate only on the surface of the lattice, and move only between faces that are exposed. 

A reconfiguration routine considers both initial $\Gamma_0$ and target $\Gamma_f$ configurations. A region of space can be classified as part of assembly or disassembly if it is modified, an intersection, or void otherwise. Formally, given $\Gamma_0$ and $\Gamma_f$:

$$\text{type}[\vec{r}] = 
\begin{cases} 
\text{Assembly}&\text{if }\Gamma_f(r)-\Gamma_0(r) = 1,\\
\text{Disassembly}&\text{if  } \Gamma_f(r)-\Gamma_0(r) = -1, \\
\text{Intersection}&\text{if }\Gamma_f(r)=\Gamma_0(r) = 1, \\
\text{Void}&\text{otherwise} \end{cases}$$

Here we only consider problems for which intersections exist. Such a requirement considerably reduces the complexity of the problem by limiting the space in which robots can operate. Any reconfiguration with empty intersections can be reduced into a series of subproblems with intersecting regions, which can be solved by the methods presented here. We also constrain the model to conserve total size: the size of assembly equals the size of disassembly - we do not consider voxel sources or sinks for clarity (though the experienced reader will see that minor modifications in these algorithms are required to accommodate those scenarios). 

We refer to a voxel that can be added or removed, for a given time step $t$, as a \textit{frontier}. These are defined by heuristics that can be mechanical (e.g., robotic or structural requirements for addition or removal) or computational (e.g., precomputed orderings). 

We consider many of a single type of robot, each able to perform $M$ moves. These moves include locomotion, voxel detachment, voxel attachment, giving or receiving a voxel from another robot, and staying stationary. Robots have these moves limited to the directions, tangential to the voxel face they occupy and along one of any Cartesian axial direction. 

\begin{figure}[ht]
\includegraphics[width=0.5\textwidth]{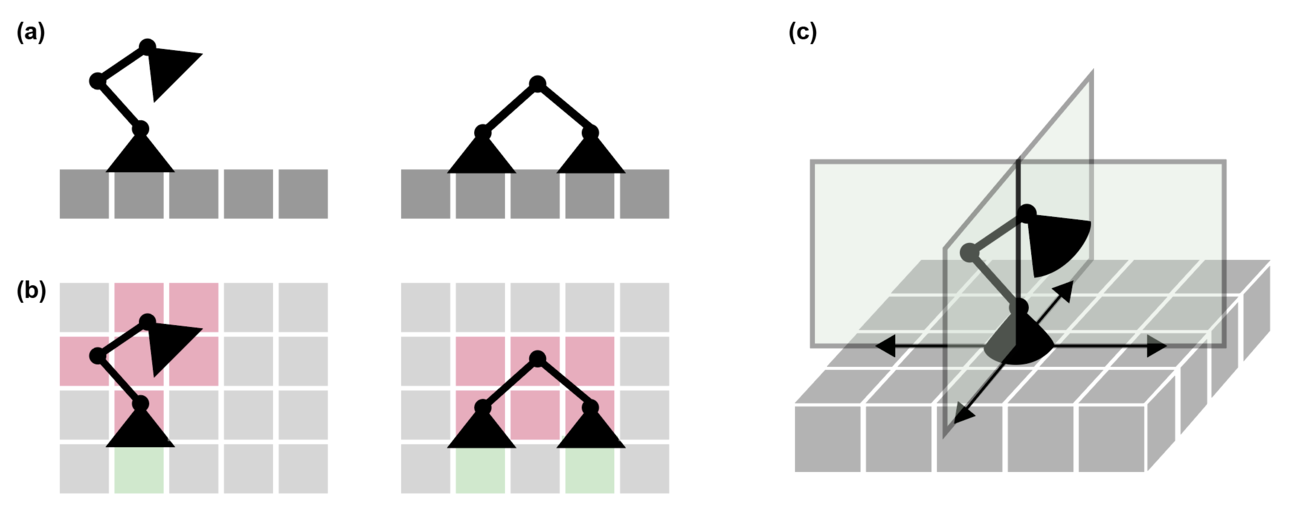}
\centering
\caption{(a) Two possible moves that a robot can perform, steady and single step. (b) Those moves are only possible if the positions in red are free and the positions in green have voxels. (c) When moving, robots can choose from four possible directions. }
\label{fig:moves}
\end{figure}

Every move is abstracted by a number of relative positions that must be free in order for the move to be possible (Figure \ref{fig:moves}), and similarly by positions that must have voxels. A move is considered to be possible only if those two conditions hold. Note that a robot that is stationary also has space requirements.

We refer to the set of moves executed by each robot at a given time step $t$ as a \textit{move step}. We refer to the time-ordered list of all move steps as a \textit{move plan}, $\Pi$, which describes a reconfiguration together with an initial configuration. A configuration $\Gamma$ is said to be \textit{unreachable} or \textit{unfeasible} from $\Gamma_0$ if there's no valid $\Pi$ that is able to reach $\Gamma$ starting
from $\Gamma_0$. Note that $|\Pi| = T =$ \textit{Assembly Time} is the number of time steps required for the reconfiguration.

\section{Algorithm Selection}

The objective function pursued here is to guarantee the reconfiguration $\Gamma_0 \rightarrow \Gamma_f$ while possibly minimizing total traveled distance, ensuring fault-tolerance, and/or proceeding in structurally optimal ways. We present two substantially different approaches, below.

We consider that centralized algorithms focus on building the move plan as a precomputed optimal solution, whereas distributed algorithms never have a single entity that holds or processes the complete move plan. For centralized algorithms, we will describe different implementations of a state space search that minimizes assembly time. 

Distributed and swarm algorithms have weak guarantees of optimality, and are never able to evaluate or parameterize a full state. In the extreme case of a cellular automaton, a finite set of rules describe how robots operate based on localized states. These algorithms are unaffected by the size of the reconfiguration. We describe an cellular automata-like algorithm in which robots use a precomputed map of gradient fields to make local decisions.

\subsection{Centralized: State Space Search}

In the State Space Search formulation, we consider the set $S$ of all possible configurations achievable from $\Gamma_0$. If the space isn't bounded, $|S|$ is infinite. In the state exploration graph, there's an edge between nodes corresponding to $\Gamma_i$ and $\Gamma_j$ if and only if there is a valid move step that allows the transformation $\Gamma_i \rightarrow \Gamma_j$. The starting node corresponds to an initial configuration $\Gamma_0$ and the target node is a goal configuration $\Gamma_f$.

Choosing to use heuristics, we use $A^* search$  to explore the graph. This formulation allows encoding and minimization of a number of different objectives: weights can be added to move steps to avoid robotic actions that are costly, heuristic costs can be added to configurations that have undesirable mechanical properties, and/or probabilities of failures can be parameterized to minimize the chances of robotic mistakes. A resulting move plan $\Pi$ can be considered to holistically consider all such parameters. In the implementation described here, we only consider heuristic costs that guide the search, but let edge costs be one in order to achieve minimum assembly time. 

If a robot can stay in a voxel in $D$ different ways, and if we bound the space by a cube of size $n$, the size of the graph is:
$$ |V| = O( (\text{D}+1)^{n^3}) $$
And there are
$$ |E|/|V| =O(M^R) $$
outgoing edges from each node. It is hence crucial that we make good usage of time and memory space. 

A number of heuristics are used to prune the exploration tree. Trivially, configurations in which voxels are being added to regions that are not assembly, or which has robots stepping out of bounds, are promptly pruned in the exploration. 

An initial heuristic approach for $A^*$ involved moving the center of mass of voxels that are removed to the center of mass of the assembly region, e.g. minimizing such distances. A later approach involved orienteering robots to gradient fields as in the distributed case. The fastest and most successful heuristic so far, however, involves computing the min-cost of a dynamic bipartite matching. 

\begin{figure}[ht]
\centering
\includegraphics[width=0.5\textwidth]{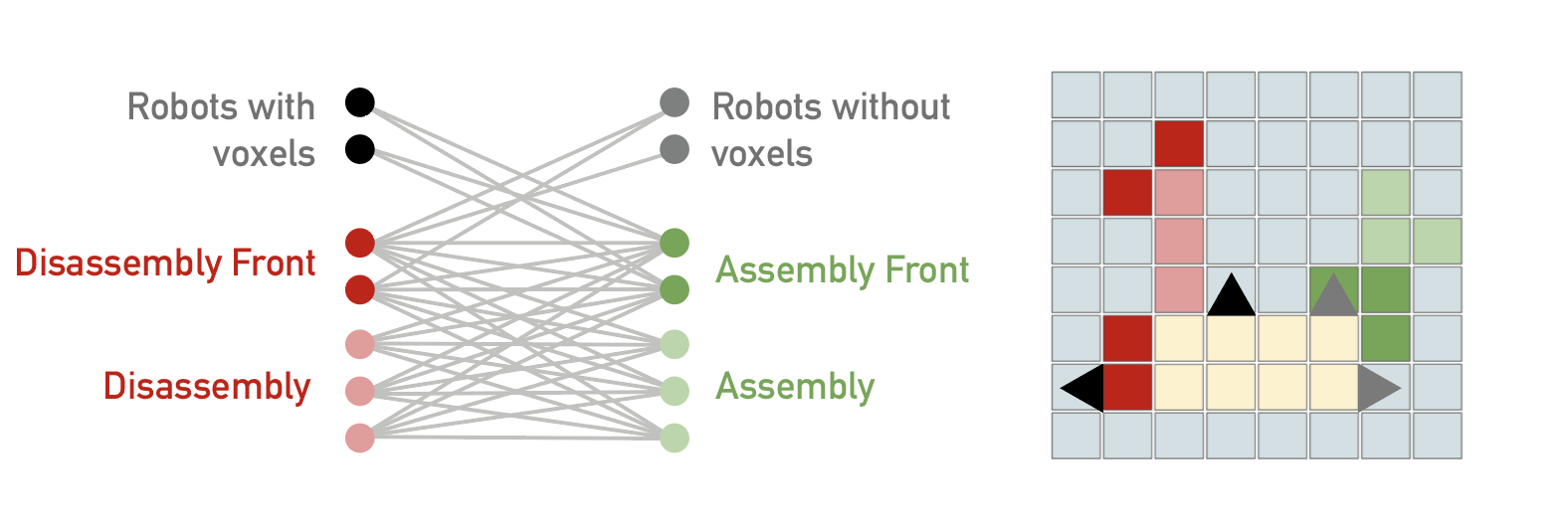}
\caption{A bipartite graph as used to compute our heuristic, and a corresponding possible configuration.}
\label{fig:bipartite_graph}
\end{figure}

We describe this heuristic now (Figure 3). We denote the sets of robots without and with voxels $R^0$ and $R^*$, respectively. $D$ and $A$ denote the set of assembly and disassembly voxels, and $DF$ and $AF$ their respective fronts. Given a configuration, we build the corresponding bipartite graph $B:(U, V)$, where $U = R^* \cup D$ and $V = R^0 \cup A$. There's an edge connecting every node in $R^*$ and $AF$, $R^0$ and $DF$, and $A$ and $D$. Edge weights denote the (surface-based) distance between their corresponding positions. The value of our heuristic is then the min-cost max-matching of $B$.  

As voxels are removed from the disassembly front, their corresponding nodes are removed from the bipartite graph. The same is done for voxels added in the assembly front. In the target configuration, the bipartite graph has the smallest number of nodes and no edges, and hence the min-cost max-matching reaches a minimum of zero. Trying to minimize this heuristic hence serves as a guide for our state exploration. It's important to note that a search might go through local minima such that the final solution might not be the optimal one if we don't explore an additional number of configurations in the $A^*$ priority queue. Increasing the number of additional states we explore is required to finding an optimal solution, but also makes our algorithm computationally unfeasible as the number of robots increase. We hence upper bound this number for computational purposes, using state space search as an approximation algorithm for increasing problem size and number of robots.

For efficiency in implementation, we only store in program memory the initial and target configurations, as well as a small buffer with the top-scored configurations. In our implementation, an initial bipartite graph is built and dynamically updated as we explore the states. As we explore the graph, we save only the hash values of the explored configurations, and for edges we only keep the corresponding move steps. If we need to explore a configuration $\Gamma$ that is not available in memory, we backtrack through the hash values to reconstruct the move plan that generates $\Gamma$ through $\Gamma_0$. 

%\vspace{30 mm}

\subsection{Distributed: Gradient Fields}
In this approach, we assume that there's no centralized authority able to compute the full solution online. Rather, robots have a precomputed map of the reconfiguration, and use both this map and local configuration to come up with a decision. 

Computing these levels draws a comparison with physics: we assign positive charges to region of space in which voxels are to be added, and negative charges to regions of voxels to be removed. The field is computed over the surface of the structure (in opposition to the full 3D space), and the resulting field lines describe paths that robots are able to follow. 

Robots can be visualized as charges following those lines. Interaction with other robots and the presence of local potential deadlocks induce robots to flip charges and navigate against the fields, adding to the complex behavior of assembly. The resulting movement searches for frontier voxels to be removed or frontier regions that can have a voxel being added. When the potential gradient doesn't point to a particular direction, the robot picks a direction at random. Finally, robots have a short memory buffer that optimizes for exploration not to continue to regions that have been visited. The program halts when no robots carry voxels, and robots following lines towards disassembly can't find any voxels to be removed.

We implement this process in our system, assuming that inert voxels are unable to store information. A first step is needed where there is a global compiler that turns the voxel map into a discrete gradient field where each voxel face now has a value that indicates how close it is to the assembly and disassembly front. In the intersecting regions of the reconfiguration, a simple equation describes how the potential field is calculated:
$$ \Phi(v) = \sum_{c \in \text{assembly}}\frac{1}{d(v,c)}\;\; - \sum_{c \in \text{disassembly}} \frac{1}{d(v,c)}$$ 
where $d(v, c)$ describes the breadth-first-search distance from voxel $c$ to face $v$. The potential is considerably trickier in the assembly and disassembly regions, where the ordering of voxel additions and removals is crucial. We use different methods for computing those depending on the geometry of the reconfiguring shapes. Those methods include: ordering the potential in a frontier-based breadth-first-search of voxels from the intersecting regions; ordering from inside-out based on the geometrical skeleton of the structure; using values drawn from functional representations of the geometries, etc. We found that different methods of calculating those allow for different reconfigurations to be performed, while causing potential field deadlocks in other ones. Further work will explore finding one such heuristic able to generalize to a number of different complex geometries, or designing an algorithm to decide which heuristic to use given the geometry problem.

\begin{figure}[ht]
\centering
\includegraphics[width=0.45\textwidth]{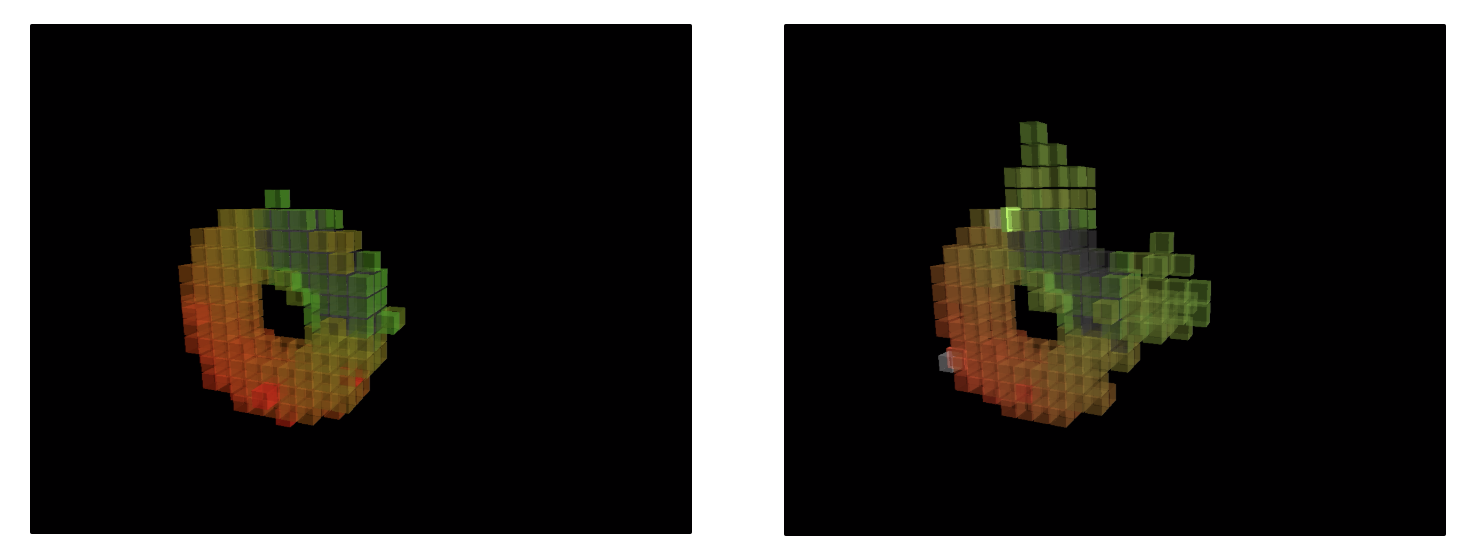}
\caption{The distributed algorithm makes usage of complex gradient fields to guide robotic movement and action. Robots use local field information and can be compared charged particles moving under a force field. }
\label{fig:gradient}
\end{figure}

A critical assumption to the correctness of our distributed algorithm is the capacity of robots to hand voxels to one another.  It is often the case in our algorithm that robots carrying voxels and robots free of voxels traverse the same field lines in opposite directions; voxel exchange is critical for conflict resolution in this scenario, and when the number of robots increases, a bucket-brigade behavior starts to appear, which increases the voxel transport efficiency. For more complex conflict resolution and bottlenecks, we define a right-of-pass based on each robot's unique number identifier to break the symmetry. Robots that can't move stand still until they are given right-of-pass.

\section{Implementation and Testing}

The algorithms were implemented in software using Java. A class \texttt{Configuration} abstracts the state of voxels and robots, and classes \texttt{MoveStep} and \texttt{MovePlan} abstract changes in these states. 
Configurations can be encoded by either three dimensional bit matrices or by functional representations (frep) that describe voxel positions. This voxel map is used together with a list of robot positions to generate configurations. 

A simulator interface is used to generate the behavior of robots under specific algorithms and conditions. This interface allow us to assign probabilities of failure to robots and voxels. A simulation outputs a file that discretely describes all the steps (including potential failures) of the robots until assembly is completed. We analyze the reconfiguration process through a visualizer that can be invoked on this file %(Figure \ref{fig:visualizer}).

\begin{figure}[ht]
\includegraphics[width=0.4\textwidth]{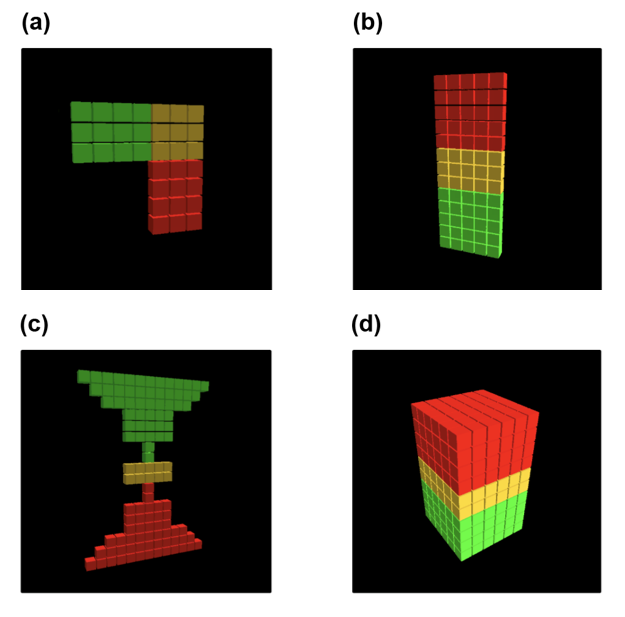}
\centering
\caption{Reconfigurations used for comparing state space search and distributed algorithms. Yellow voxels indicate intersecting regions between the initial and target configurations. Green and red voxels indicate the assembly and disassembly regions, respectively. }
\label{fig:tests}
\end{figure}

We limit our tests to comparing state space search with distributed automata. We test our algorithms in 7 different scenarios, using 4 different shapes (Figure \ref{fig:tests}) with varying number of robots. The shapes and number of moved voxels are as follows: a) The rotation of a rectangle, 12 voxels. b) Linear translation of rectangle, 25 voxels. c) 2D sand timer. 50 voxels. This shape is used to test bottlenecks and funneling effects in structures. d) Translation of a large cube, 144 voxels.

We also impose different restrictions to robotic movement. Degrees of freedom of 2 are such that robots are limited to a single plane of movement, without depth. For example, we might impose this restriction to a robot in shape (a), so that it can only move in the border of the structure. This restriction is equivalent to requiring that the robot in Fig. 2 (c) be constrained to a single plane to move. This contrasts with degrees of freedom of value 3, in which both planes (i.e. all four directions) can be chosen, and robots can move over the whole surface of the structure. 

\begin{figure}[ht!]
\centering
\includegraphics[width=0.45\textwidth]{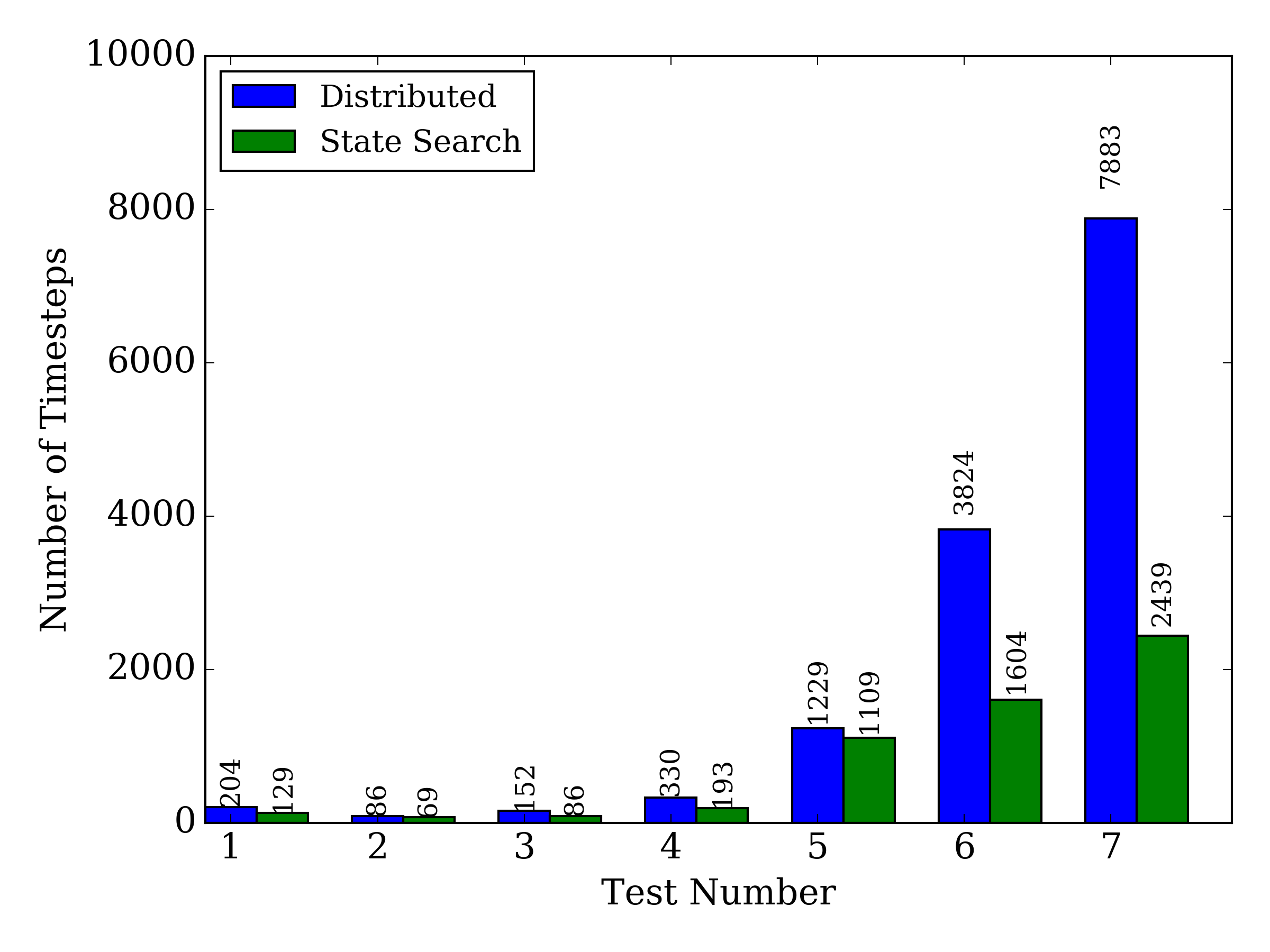}
\caption{Comparison between the distributed and state search total elapsed time for the test cases.}
\label{fig:TestsTime}
\end{figure}

Centralized algorithms have strong guarantees of optimality on structure assembly or total assembly time. This phenomena is apparent in figures \ref{fig:TestsTime} and \ref{fig:TestsSteps}, where the state search algorithm outperforms the distributed algorithm in terms of minimizing both the travel distance and assembly time, hence making the system less prone to failure. It is also apparent that the difference in performance between both methods also increases when the number of voxels, robots or degree of freedom increases. The state search total assembly time is 60\% less than the distributed algorithm for test 1 and 230\% less for test 7. 

\begin{figure}[ht]
\centering
\includegraphics[width=0.45\textwidth]{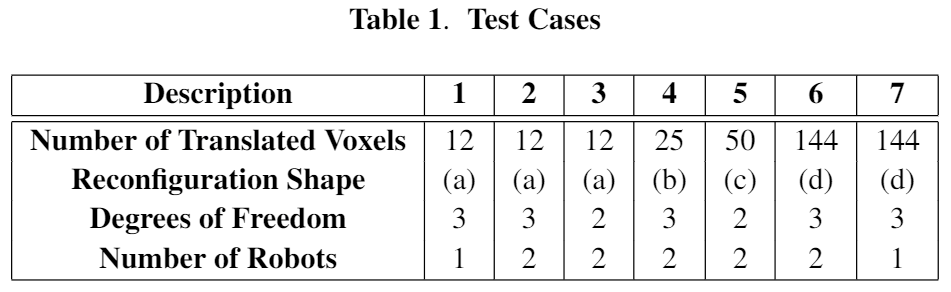}
\label{fig:testsDescription}
\end{figure}

\begin{figure}[H]
\centering
\includegraphics[width=0.45\textwidth]{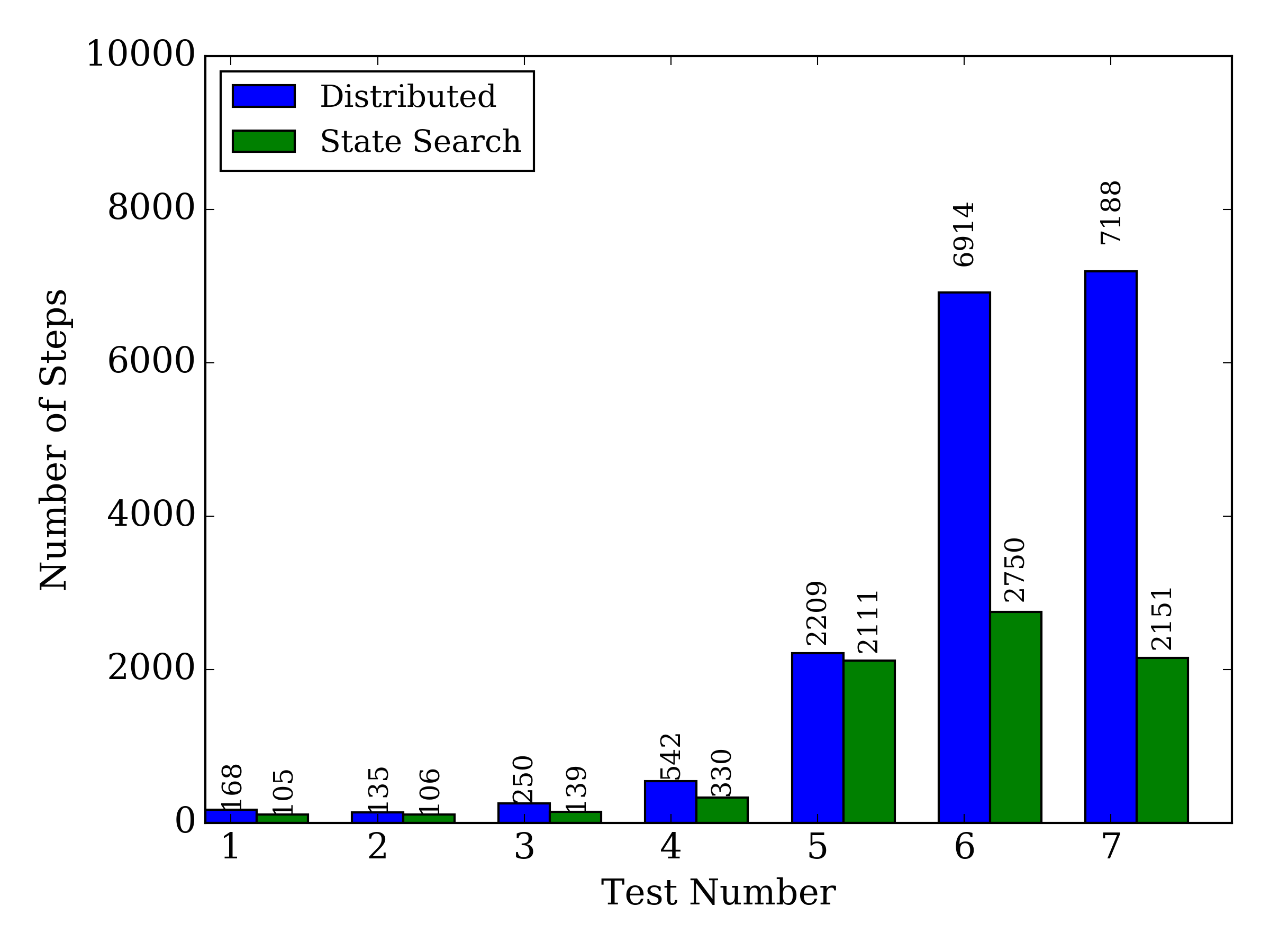}
\caption{Comparison between the distributed and state search total distance traveled for the test cases.}
\label{fig:TestsSteps}
\end{figure}

\begin{figure}[t]
\centering
\includegraphics[width=0.45\textwidth]{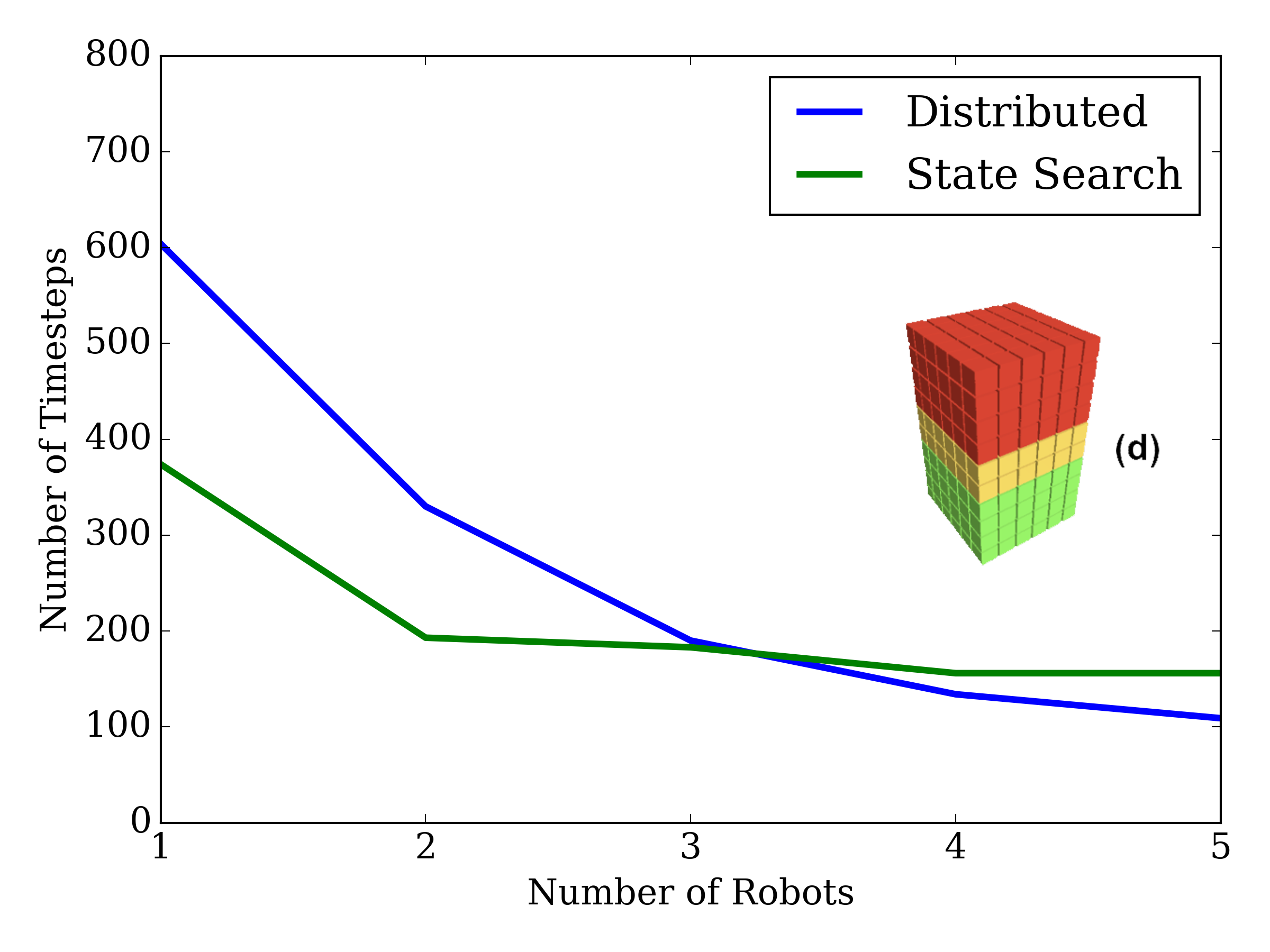}
\caption{Comparison of total elapsed time between the distributed and the state algorithm on shape (d) for 1 to 5 randomly placed robots.}
\label{fig:DistvsStateTime}
\end{figure}

\begin{figure}[t]
\centering
\includegraphics[width=0.45\textwidth]{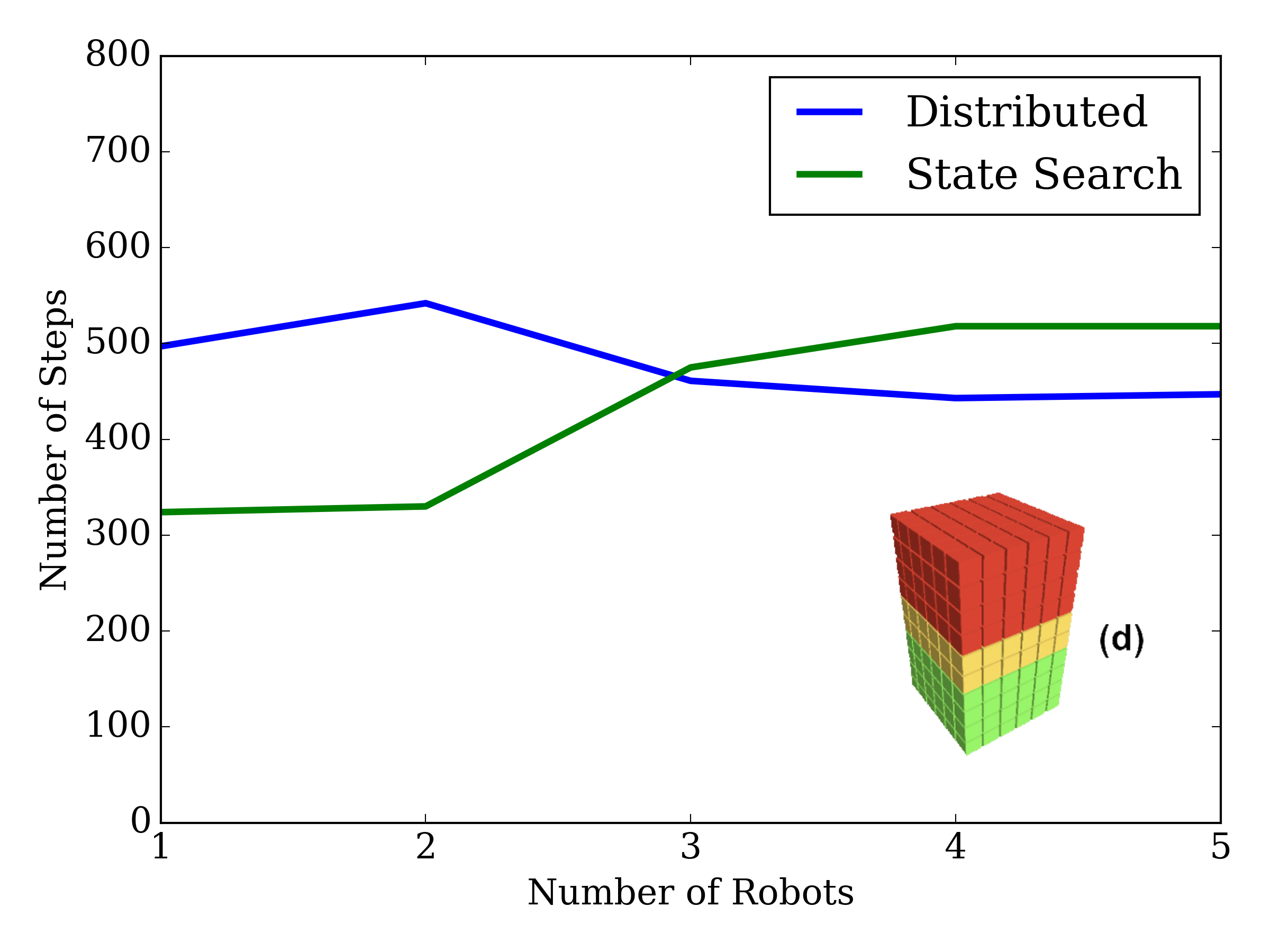}
\caption{Comparison of total traveled distance between the distributed and the state algorithm on shape (d) for 1 to 5 randomly placed robots.}
\label{fig:DistvsStateStep}
\end{figure}

\begin{figure}[ht]
\centering
\includegraphics[width=0.45\textwidth]{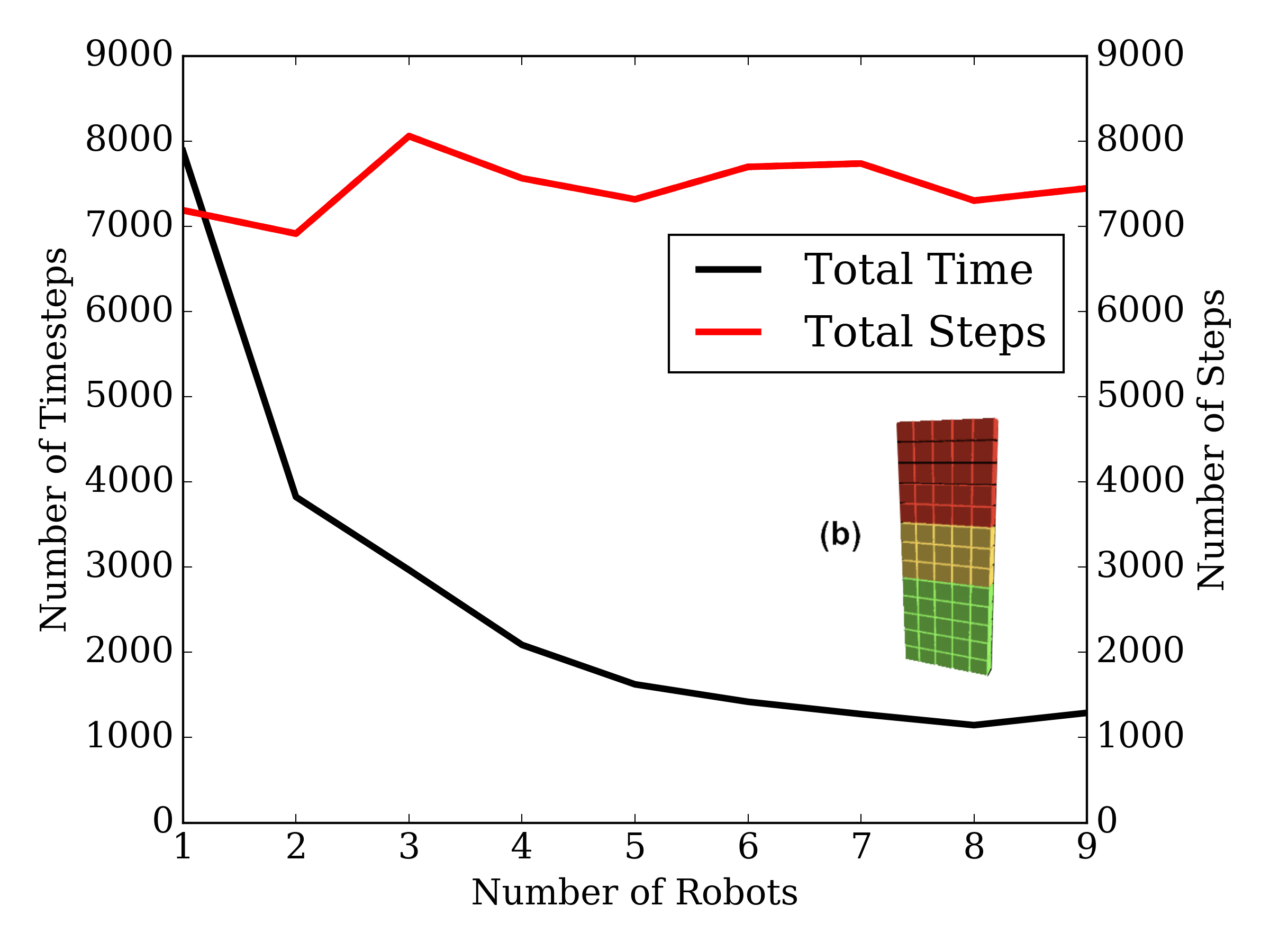}
\caption{Total assembly time for the distributed algorithm on shape (b) for 1 to 8 randomly placed robots showing how the total time drastically decreases with time.}
\label{fig:RobotvsTimevsStep}
\end{figure}

Nevertheless, even if centralized algorithms often give us optimal solutions, since these solutions are precomputed, any robotic or structural failure will require further expensive recomputation. It is important to notice that when the size of the problem increases, fully centralized approaches become unfeasible and we turn to approximations for state space search. Moreover, we don't allow voxel exchange in our centralized approach, yet this is a critical assumption in the distributed case. These two factors explain figures \ref{fig:DistvsStateTime} and \ref{fig:DistvsStateStep} where, when the number of robots increases in relation to the number of voxels, the solutions that both solve complex collisions and perform better that the distribution algorithm become rare.

Distributed algorithms are unaffected by the size of the reconfiguration, and are generally oblivious to global situations such as robotic or structural failures. Figure \ref{fig:RobotvsTimevsStep} shows how the total assembly time for shape (b) drastically decreases when the number of robots increases while the total travel distance remains the same. However, when the density of robots increases, the systems becomes saturated and more collisions occur between the robots, increasing both the assembly time and total travelled distance. This phenomena counteracts the savings of adding extra robots, stabilizing the total assembly time in relation of number of robots.

% The implementation of $A^* search$ described above does not assume edge weights, but those can easily be drawn from costs of robotic moves.

\section{Conclusion}
\begin{figure}[h]
\centering
\includegraphics[width=0.45\textwidth]{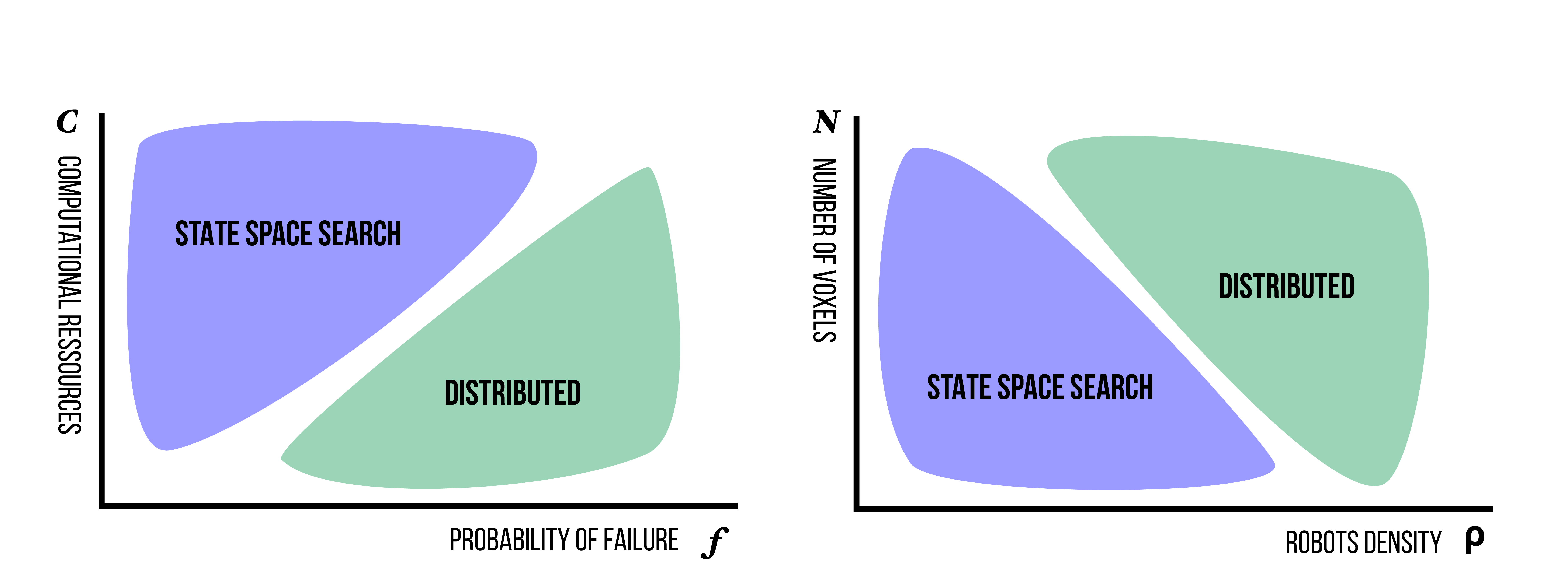}
\caption{Conceptual Ashby-style plot of the optimum performance of each algorithm in relation to four problem parameters: computational resources, probability of failure, total number of voxels and robot density.}
\label{fig:ashby2}
\end{figure}

From the conducted study, we were able to draw some conclusions regarding the feasibility and optimally of using a number of algorithms for the reconfiguration of discretely assembled structures. In general, it is important to note that the effectiveness of each algorithm is heavily dependent on the objective function, and there isn't a single absolute approach that would have the optimum performance under all conditions (Figure \ref{fig:ashby2}). For instance, failure rate is an important factor when trying to chose a control algorithm. For centralized systems, the precomputed solutions have to be reevaluated more frequently. A windowed precomputation minimizes risk of losing too much of a precomputed solution, but might result in solutions that are far from optimal.  A high number of robots compared to number of voxels implies greater amounts of possible system deadlocks, and might require centralized approaches to find solutions. Centralized approaches were shown to perform considerably better than distributed automata in small reconfigurations. However, an increasing number of robots and voxels requires centralized approaches to be bounded and approximated, which can result in worse results from a greedy distributed approach. Still, even for those cases centralized approaches can be used for finding solutions that require holistic understanding of the system (e.g., structural requirements), compared to distributed. Further studies need to be done in mixed algorithms in which large scale distribution is combined with localized centralized computations. 

In conclusion, the paper pointed out and discussed the limitations of each algorithm in relation to different problem conditions. Finding an optimal algorithm is itself a multi-objective optimization problem for different geometries, robot densities and problem sizes, and we addressed variations of these situations with representative algorithms and test cases.

\section*{Acknowledgments}

This research was supported by the Automated Reconfigurable Mission Adaptive Digital Assembly Systems (ARMADAS) project through NASA STMD Game Changing Development (GCD) Program, the NASA Space Technology Research Fellowship (NSTRF) Grant \#NNX14AM40H, the NASA International Internship program, the MIT International Science and Technology Initiative (MISTI), and the Center for Bits and Atoms (CBA) consortia funding. 

%\ifCLASSOPTIONcaptionsoff
 % \newpage
%\fi



%\bibliographystyle{IEEEtran}
% argument is ythe BibTeX string definitions and bibliography database(s)

\bibliography{bib1.bib}
\bibliographystyle{IEEEtran}

% ythe own custom macro containing the \includegraphics command to make things
% \begin{IEEEbiography}[{\includegraphics[width=1in,height=1.25in,clip,keepaspectratio]{mshell}}]{Michael Shell}

%%%%%%%%%%%%%%%%%%%%%%%%%%%%%%%%%%%%%%%%%%
\newpage
\thebiography
%% This biostyle allows you to insert your photo size 1in X 1.25in

\begin{biographywithpic}
{Allan Costa}{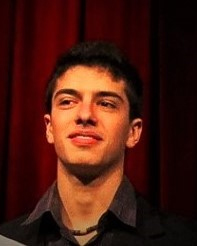}
is an undergraduate at the Massachusetts Institute of Technology, where he is pursuing his bachelor's degree in computer science \& engineering, and physics. He has worked with MIT's Center for Bits and Atoms and NASA Ames Research Center, and is currently researching abroad at ETH Zurich.
\end{biographywithpic}

\begin{biographywithpic}
{Amira Abdel-Rahman}{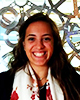}
is a graduate student researcher at the Massachusetts Institute of Technology's Center for Bits and Atoms, where she is pursuing her Ph.D. on generative computational tools for discrete structures and materials. She holds a Master's degree from Harvard Graduate School of Design.
\end{biographywithpic}

\begin{biographywithpic}
{Benjamin Jenett}{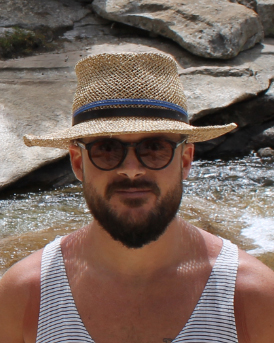}
is a graduate student researcher at the Massachusetts Institute of Technology's Center for Bits and Atoms, where he is pursuing his Ph.D. on automated assembly for large aerospace structures. He was formerly a NASA Space Technology Research Fellow and is currently in the NASA Pathways program. 
\end{biographywithpic}

\begin{biographywithpic}
{Irina Kostitsyna}{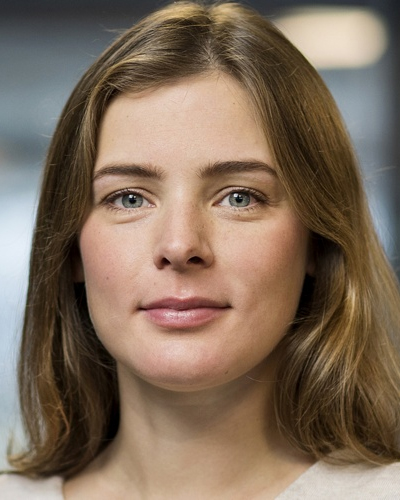}
is an assistant professor in the Applied Geometric Algorithms group in the Department of Mathematics and Computer Science. Her research interests lie in the field of computational geometry, both in its theoretical and applied aspects. She holds a Ph.D. from Stony Brook University. 
\end{biographywithpic}

\begin{biographywithpic}
{Neil Gershenfeld}{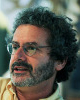}
is the Director of MIT’s Center for Bits and Atoms. Dr. Gershenfeld has a BA in Physics with High Honors from Swarthmore College, a Ph.D. in Applied Physics from Cornell University, honorary doctorates from Swarthmore College, Strathclyde University and the University of Antwerp. 
\end{biographywithpic}

\begin{biographywithpic}
{Kenneth Cheung}{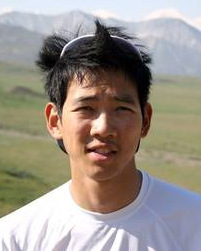}
received his Ph.D. from the Center for Bits and Atoms at the Massachusetts Institute of Technology. He helps to run the ARC Coded Structures Laboratory (CSL), which conducts research on the application of building block based materials and algorithms to aeronautical and space systems. %As a member the NASA ARC Intelligent Systems Division and affiliate of the office of the Center Chief Technologist, he serves as a technical lead on advanced materials and manufacturing.
\end{biographywithpic}

\end{document}